\documentclass[12pt,preprint]{aastex}
\begin{document}
\title{CONSTRAINING THE DARK ENERGY EQUATION OF STATE WITH COSMIC VOIDS}
\author{Jounghun Lee and Daeseong Park}
\affil{Department of Physics and Astronomy, FPRD, Seoul National University, 
Seoul 151-747, Korea} 
\email{jounghun@snu.ac.kr}
\begin{abstract}
Our universe is observed to be accelerating due to the dominant dark energy
with negative pressure. The dark energy equation of state ($w$) holds a key 
to understanding the ultimate fate of the universe. The cosmic voids behave 
like bubbles in the universe so that their shapes must be quite sensitive to 
the background cosmology. Assuming a flat universe and using the priors on 
the matter density parameter ($\Omega_{m}$) and the dimensionless Hubble 
parameter ($h$), we demonstrate analytically that the ellipticity evolution 
of cosmic voids may be a sensitive probe of the dark energy equation of state. 
We also discuss the parameter degeneracy between $w$ and $\Omega_{m}$. 
\end{abstract}


Recent observations have revealed that our universe is flat and in a phase of 
acceleration \citep{rie-etal98,per-etal99,spe-etal03}. It implies that some 
mysterious dark energy fills dominantly the universe at present epoch, 
exerting anti-gravity. The nature of this mysterious dark energy which holds 
a key to understanding the ultimate fate of the universe is often specified 
by its equation of state, i.e., the ratio of its pressure to density: 
$w\equiv P_{de}/\rho_{de}$. The anti-gravity of the dark energy corresponds 
to the negative value of $w$. The simplest candidate for the dark energy is 
the vacuum energy ($\Lambda$) with $w=-1$ that is constant at all times 
\citep{ein17}. Although all current data are consistent with the vacuum 
energy model \citep[e.g.,][]{wan-teg04,jas-etal04,per05,guz-etal08}, the 
notorious failure of the theoretical estimate of the vacuum energy density  
\citep[see][for a review]{car-etal92} has led a dynamic dark energy model to 
emerge as an alternative. In this dynamic dark energy models which is 
often collectively called quintessence, the dark energy is described as a 
slowly rolling scalar field with time-varying equation of state in the 
range of $-1<w<0$ \citep{cal-etal98}. 

The following observables have so far been suggested to discriminate the 
dark energy models: the luminosity-distance measure of type Ia supernova 
\citep{rie-etal04,rie-etal07,dav-etal07,kow-etal08}; the abundance of galaxy 
clusters as a function of mass \citep{wan-ste98,hai-etal01,wel-etal02}, 
the baryonic acoustic oscillations in the galaxy power spectrum 
\citep{bla-gla03,hu-hai03,coo04,seo-eis05}, and the weak gravitational 
lensing effect \citep{hu99,hut01,tak-jai04,son-kno04}. True as it is that 
these observables can constrain powerfully  the value of $w$, it is still 
quite necessary and important to find out as many different observables 
as possible for consistency tests. 

Another possible observable as a dark energy constraint may be the shapes 
of the cosmic voids. As the voids behave like bubbles due to their extremely 
low densities, their shapes determined by the spatial distribution of the 
void galaxies tend to change sensitively according to the competition between 
the tidal distortion and the gravitational rarefaction effect. Therefore, 
the shape evolution of the voids must depend sensitively on the 
background cosmology. In this Letter we study the ellipticity evolution of 
cosmic voids in the QCDM (quintessence + cold dark matter) model with the 
help of the analytic formalism developed by \citet{par-lee07} and explore 
the possibility of using it as a complimentary probe of the dark energy 
equation of state. 
 
According to \citet{par-lee07}, the shape of a void region is related 
to the eigenvalues of the local tidal shear tensor as 
\begin{eqnarray}
\label{eqn:lamu1}
\lambda_{1}(\mu,\nu) &=& \frac{1 + (\delta_{v}- 2)\nu^{2} + 
\mu^{2}}{(\mu^{2} + \nu^{2} + 1)},\\
\label{eqn:lamu2} 
\lambda_{2}(\mu,\nu) &=& \frac{1 + (\delta_{v}- 2)\mu^{2} + \nu^{2}}
{(\mu^{2} + \nu^{2} + 1)},
\end{eqnarray}
where $\{\lambda_{i}\}_{i=1}^{3}$ (with $\lambda_{1}>\lambda_{2}>\lambda_{3}$) 
are the three eigenvalues of the local tidal field smoothed on void scale, 
$\delta_{\rm v}$ is the density contrast threshold for the formation 
of a void: $\delta_{\rm v}=\sum_{i=1}^{3}\lambda_{i}$, and $\{\mu,\nu\}$ 
(with $\nu <\mu$) represents a set of the two parameters that quantify the 
anisotropic distribution of the void galaxies. They defined the void 
ellipticity as $\varepsilon\equiv 1 -\nu$ and evaluated its probability 
density distribution as 
\begin{eqnarray} 
p(1-\varepsilon;z) &=& p(\nu;z,R_{L}) =
\int_{\nu}^{1} p[\mu,\nu|\delta =\delta_{\rm v};\sigma(z,R_{L})]d\mu \cr
&=&\frac{3375\sqrt{2}}{\sqrt{10\pi}\sigma^{5}(z,R_{L})}
\exp\left[-\frac{5\delta^{2}_{\rm v}}{2\sigma^{2}(z,R_{L})} + 
\frac{15\delta_{\rm v}(\lambda_{1}+\lambda_{2})}{2\sigma^{2}(z,R_{L})}
\right]\cr &&\times\exp\left[-\frac{15(\lambda^{2}_{1}+\lambda_{1}
\lambda_{2}+\lambda^{2}_{2})}{2\sigma^{2}(z,R_{L})}\right]
(2\lambda_{1}+\lambda_{2}-\delta_{\rm v})\cr
&&\times(\lambda_{1}-\lambda_{2})(\lambda_{1}+2\lambda_{2}-\delta_{\rm v})
\frac{4(\delta_{\rm v}-3)^2\mu\nu}{(\mu^{2}+\nu^{2}+1)^{3}}.
\label{eqn:con}
\end{eqnarray}
Here, $\sigma(z,R_{L}))\equiv D^{2}(z)\int_{-\infty}^{\infty}\Delta^{2}(k)
W^{2}(kR_{L})d\ln k$ is the linear rms fluctuation of the matter density 
field smoothed on a Lagrangian void scale of $R_{L}$ at redshift $z$  where 
$D(z)$ is the linear growth factor, $W(kR_L)$ is a top-hat window function, 
and $\Delta^{2}(k)$ is the dimensionless linear power spectrum. Throughout 
this study, we adopt the linear power spectrum of the cold dark matter 
cosmology (CDM) that does not depend explicitly on $w$ \citep{bar-etal86}. 

Equation (\ref{eqn:con}) was originally derived under the assumption of 
a $\Lambda$CDM model ($w=-1$). We propose here that it also holds good 
for the case of a QCDM (quintessence+CDM) model where the dark energy 
equation of state changes with time as $w(z)=w_{0}+w_{a}z/(1+z)$  
\citep{cp01,lin03} where $w_{0}$ is the value of $w$ at present epoch and 
$w_{a}$ quantifies how the dark energy equation of state changes with time.  
Then, we employ the following approximation formula for the linear growth 
factor, $D(z)$, for a QCDM model \citep{bas03,per05}:  
\begin{equation}
D(z)=\frac{5\Omega_{m}}{2(z+1)}
\left[\Omega^{\alpha}_{m}-\Omega_{Q} + 
\left(1+\frac{\Omega_{m}}{2}\right)
\left(1+{\cal A}\Omega_{Q}\right)\right]^{-1}.
\label{eqn:D}
\end{equation}
where   
\begin{eqnarray}
\label{eqn:e2}
E^{2}(z)&=&\Omega_{m}(1+z)^{3}+\Omega_{Q}(1+z)^{-f(z)},\\
\label{eqn:fz}
f(z) &=& -3(1+w_{0})-\frac{3w_{a}}{2\ln(1+z)},\\
\label{eqn:alp}
\alpha &=&\frac{3}{5-2/(1-w)}+\frac{3}{125}
\frac{(1-w)(1-3w/2)}{(1-6w/5)^{3}}[1-\Omega_{m}],\\
\label{eqn:cal}
{\cal A}&=&-\frac{0.28}{w+0.08}-0.3.
\end{eqnarray}
The CDM density parameter $\Omega_{m}$ and the dark energy density parameter 
$\Omega_{Q}$ evolve with $z$ respectively as 
\begin{equation}
\Omega_{m}(z)=\frac{\Omega_{m0}(1+z)^{3}}{E^{2}(z)}, \quad
\Omega_{Q}(z)=\frac{\Omega_{Q0}}{E^{2}(z)(1+z)^{f(z)}}, 
\label{eqn:ome} 
\end{equation}
where $\Omega_{m0}$ and $\Omega_{Q0}$ represent the present values.
Equation (\ref{eqn:con}) implies that the mean ellipticity of voids  
decreases with $z$. A key question is how the rate of the decrease changes 
with the dark energy equation of state. Since most of the recent observations 
indicate that the dark energy equation of state at present epoch is consistent 
with $w=-1$ \citep[e.g., see][and references therein]{guz-etal08} 
we focus on how the mean void ellipticity depends on the value of 
$w_{a}$. Even in case that $w_{0}=-1$, if $w_{a}$ is found to deviate 
from zero, it would imply the dynamic dark energy, disproving the 
simple $\Lambda$CDM model.

To explore how the void ellipticity evolution depends on $w_{a}$, we evaluate 
the mean ellipticity of voids as
$\bar{\varepsilon}(z)=\int_{0}^{1}~\varepsilon~p(\varepsilon; R_{L},z)
d\epsilon$ for different values of $w_{a}$ through equations (\ref{eqn:con})-
(\ref{eqn:ome}). The other key cosmological parameters are set at 
$\Omega_{m}=0.75$,$\Omega_{Q}=0.75$, $h=0.73$, $\sigma_{8}=0.9$ and 
$w_{0}=-1$. When the abundance of evolution of galaxy clusters is used to 
constrain the dark energy equation of state, the cluster mass is usually set 
at a certain threshold, $M_{R}$, defined as the mass within a certain comoving 
radius \citep{wan-ste98}. Likewise, we set the Lagrangian scale of a void, 
$R_{L}$ at $4h^{-1}$Mpc, which is related to the mean effective radius of 
a void as $\bar{R}_{E}=(1+\delta_{v})^{-1/3}\bar{R}_{L}/(1+z)$. The 
Lagrangian scale $R_{L}=4h^{-1}$Mpc corresponds to the mean effective 
size of a void at present epoch, $R_{E}\sim 8.5h^{-1}$Mpc. 

\begin{figure} 
\plotone{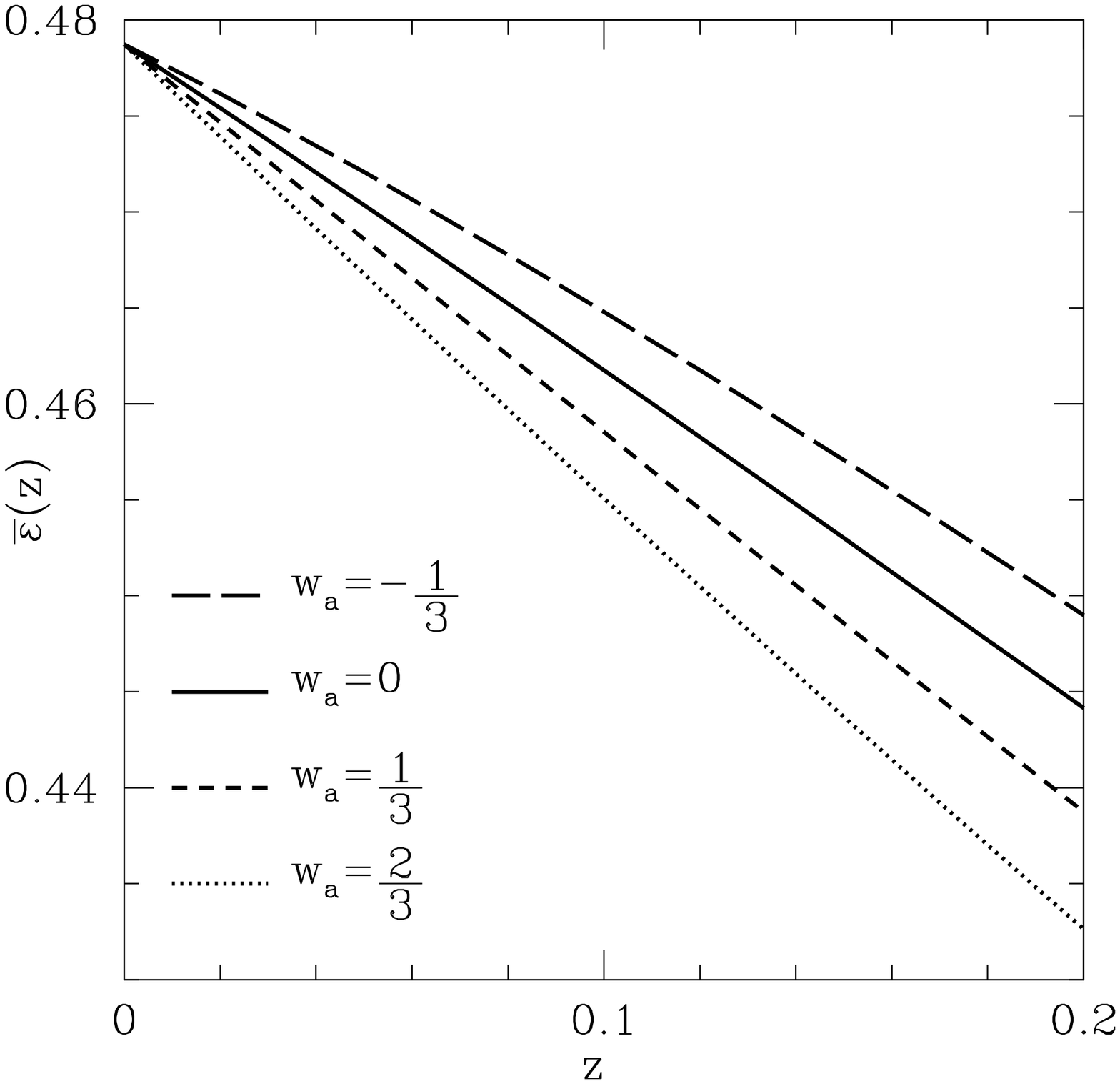}
\caption{Mean ellipticity of the voids with $R_{\rm L}=4h^{-1}$Mpc 
as a function of $z$.}
\label{fig:bare}
\end{figure}

\begin{figure} 
\plotone{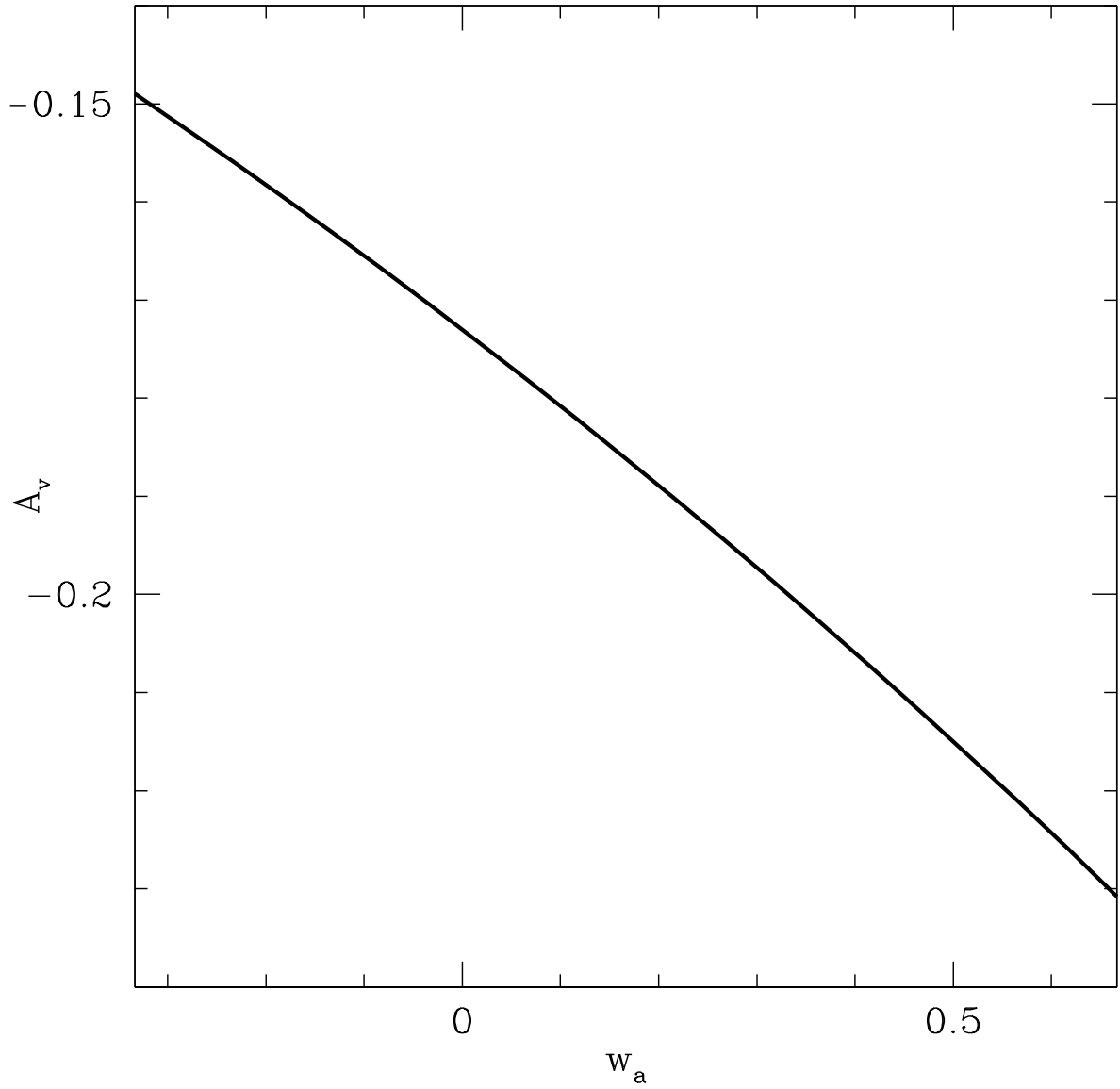}
\caption{Slope of the void ellipticity as a function of $w_{a}$.}
\label{fig:slope}
\end{figure}

Figure \ref{fig:bare} plots $\bar{\varepsilon}(z)$ for the four 
different cases: $w_{a}=-1/3,0,1/3$ and $2/3$ (long-dashed, solid, dashed, 
and dotted line, respectively). As can be seen, the higher the value of 
$w_{a}$ is, the more rapidly $\bar{\varepsilon}(z)$ decreases.  
It also suggests that $\bar{\varepsilon}(z)$ is well approximated as a linear 
function of $z$ in recent epochs ($0<z<0.2$). 
Therefore, we fit $\varepsilon(z)$ to a straight line as 
$\bar{\varepsilon}(z)\approx A_{v}z + B_{v}$. Varying the value of $w_{a}$ 
in the range of $[0,2/3]$, we compute the best-fit slope $A_{v}$. The range, 
$0\le w_{a}\le 2/3$, corresponds to the dark energy equation of state range, 
$-1\le w\le -0.9$. The result is plotted in Fig.~\ref{fig:slope}. As can be 
seen, the void ellipticity evolves more rapidly  as the value of $w_{a}$ 
increases. That is, the void ellipticity undergoes a stronger evolution when 
the anti-gravitational effect is less strong in recent epochs. Note that 
$A_{v}$ shows a noticeable $30\%$ difference as the dark energy equation of 
state changes $w$ from $-1$ to $-0.9$.

\begin{figure} 
\plotone{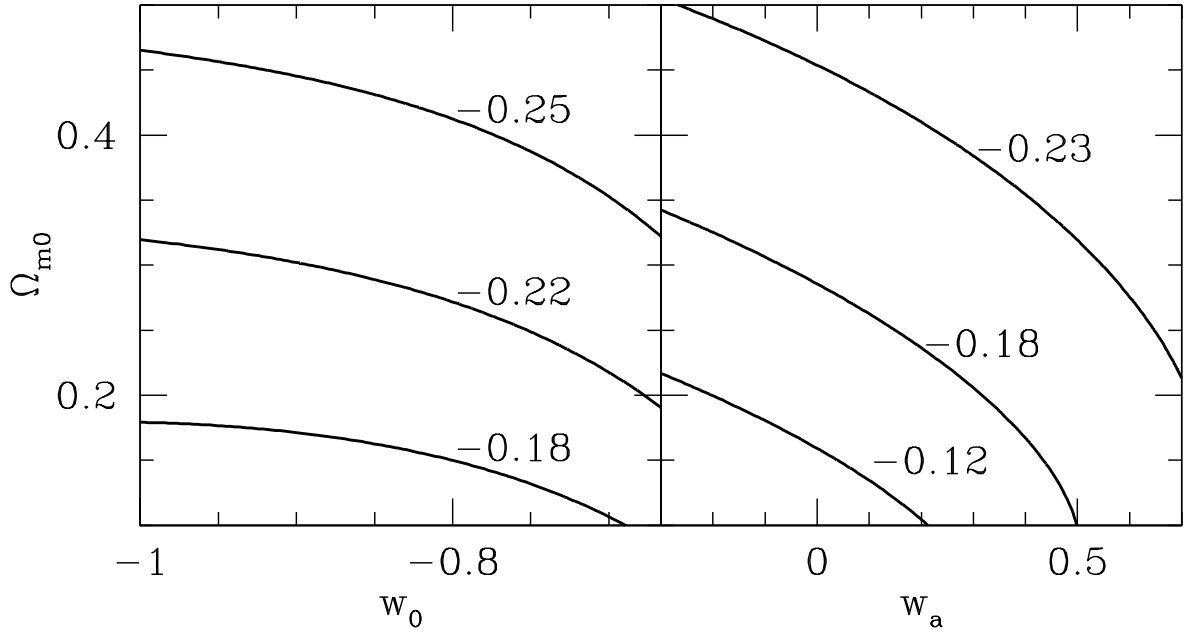}
\caption{Contours of $A_{v}$ in the $\Omega_{m0}$-$w_{0}$ (left) and 
in the $\Omega_{m}$-$w_{a}$ (right) plane.}
\label{fig:ow}
\end{figure}

We have so far neglected the parameter degeneracy between $w$ and the other 
key parameters. However, as the dependence of the void ellipticity 
distribution on the dark energy equation of state comes from its dependence 
on $\Delta^{2}(k;\Omega_{m0},\sigma_{8},h,w)$, it is naturally expected that 
there should be a strong parameter degeneracy. Here, we focus on the 
degeneracy between $\Omega_{m0}$ and $w$. First, we recompute $A_{v}$, varying 
the values of $\Omega_{m0}$ and $w_{0}$ with setting $w_{a}=1/3$.  The left 
panel of Fig. \ref{fig:ow} plots a family of the degeneracy curves in the 
$\Omega_{m0}$-$w_{0}$ plane for the three different values of $A_{v}$. 
As can be seen, there is a strong degeneracy between the two parameters. 
For a given value of $A_{v}$, the value of $w_{0}$ increases as the value 
of $\Omega_{m0}$ decreases. A similar trend is also found in the 
$\Omega_{m0}$-$w_{a}$ degeneracy curves that are plotted in the 
right panel of Fig. \ref{fig:ow} for which the value of $w_{0}$ is set 
at $-1$. It is worth noting that this degeneracy trend is orthogonal to 
that found from the cluster abundance evolution (see Fig.3 in Wang \& 
Steinhardt 1998). Thus, when combined with the cluster analysis, the void 
ellipticity analysis may be useful to break the degeneracy between 
$\Omega_{m0}$ and $w$. 

We have shown that the void ellipticity evolution is in principle a 
useful constraint of the dark energy equation of state. We have also shown 
that it provides a new degeneracy curve for $\Omega_{m0}$ and $w$. When 
combined with the cluster abundance analysis, it should be useful to break 
the degeneracy. Furthermore, unlike the mass measurement of high-$z$ clusters 
which suffers from considerable scatters, the void ellipticities are 
readily measured from the positions of the void galaxies without requiring 
any additional information.

To use our analytic tool in practice to constrain the dark energy equation of 
state, however, it will require to account for the redshift distortion effect 
since the positions of the void galaxies are measured in redshift space. 
In our companion paper (Park \& Lee 2009 in preparation), we have analyzed 
the Millennium Run Redshift-Space catalog \citep{spr-etal05} and determined 
the ellipticity distribution of the galaxy voids. From this analysis, it is 
somewhat unexpectedly found that the void ellipticity distribution measured in 
redshift space is hardly changed from the one in real space. In fact, 
this result is consistent with the recent claims of Hoyle \& Vogeley (2007) 
and that of van de Weygaert (2008, private communication) who have already 
pointed out that the redshift distortion effect has only negligible, if any, 
effect on the shapes of voids. We hope to constrain the dark energy equation 
of state by applying our theoretical tool to real observational data and 
report the result elsewhere in the near future.
 
\acknowledgments
We thank an anonymous referee for a constructive report.
J.L. am very grateful to S.Basilakos for very helpful discussion and comments.
This work is financially supported by the Korea Science and Engineering 
Foundation (KOSEF) grant funded by the Korean Government 
(MOST, NO. R01-2007-000-10246-0).

\clearpage
\clearpage
\clearpage
\end{document}